\documentstyle[prd,aps]{revtex}
\def\lb{\left(}
\def\rb{\right)}
\def\be {\begin{equation}}
\def\ee {\end{equation}  }
\def\beq{\begin{eqnarray}}
\def\eeq{\end{eqnarray}  }
\def\bi {\begin{itemize} }

\def\ei {\end{itemize}   }
\def\RE {I\kern-6pt R    }
\def\Z  {Z\kern-13pt Z   }
\def\be {\begin{equation}}
\def\ee {\end{equation}  }
\def\beq{\begin{eqnarray}}
\def\eeq{\end{eqnarray}  }

\def\eeq{\end{eqnarray}  }

\input epsf

\begin{document}
\draft

\twocolumn[\hsize\textwidth\columnwidth\hsize\csname
@twocolumnfalse\endcsname

\title{Critical Phenomena Inside Global Monopoles}

\author{Steven L. Liebling}
\address{Theoretical and Computational Studies Group\\
     Southampton College, Long Island University,
     Southampton, NY 11968}

\maketitle

\begin{abstract}
The gravitational collapse of a triplet scalar field is examined
assuming a hedgehog ansatz for the scalar field. Whereas
the seminal work by Choptuik with a single, strictly spherically
symmetric scalar field
found a discretely self-similar (DSS)
solution at criticality with echoing period $\Delta=3.44$, here a new
DSS solution is found with period $\Delta=0.46$. This new critical solution
is also observed in the presence of a symmetry breaking potential as well as
within a global monopole.
The triplet scalar field model
contains Choptuik's original model in a certain region of
parameter space, and hence
his original DSS solution is also a solution.
However, the choice of a hedgehog ansatz appears to exclude the original DSS.
\end{abstract}


\vskip2pc]

%%%%%%%%%%%%%%%%%%%%%%%%%%%%%%%%%%%%%%%%%%%%%%%%%%%%%%%%%%%%%%%%
\section{Introduction}
\label{sec:introduction}
%%%%%%%%%%%%%%%%%%%%%%%%%%%%%%%%%%%%%%%%%%%%%%%%%%%%%%%%%%%%%%%%
Quite an industry has developed around the study of nonlinear
phenomena occurring at the threshold of black hole formation
first discovered by Choptuik~\cite{matt}. A number of reviews
detail the substantial progress in understanding
these phenomena~\cite{matt_rev,carsten_rev}. By specifying a form
of initial data parameterized by some parameter $p$, numerical evolutions
can determine whether the value of $p$ is sufficient to form a black
hole. Such a solution is considered super-critical, while for small
$p$, no black hole forms and the solution is sub-critical. Separating
super-critical solutions from sub-critical solutions is the critical
solution parameterized by the critical value
$p^*$. In analogy with statistical mechanics where the black hole
mass represents the order parameter,
%cases where the critical
%solution has a finite mass are called Type~I, and
cases where infinitesimal mass black holes are formed
are called Type~II, and cases in which only finite mass black holes are formed
are called Type~I.

Here, I present a new Type~II critical solution occurring within
a model with a triplet scalar field. By assuming a hedgehog ansatz
for the triplet, I find a critical solution which is, in a strict
sense, non-spherically symmetric. Adding a potential allows the
study of
critical phenomena within a global monopole.
No Type~I critical behavior around the static monopole solutions
is observed.

As a generalization of Choptuik's original model, the triplet scalar
field contains the single scalar field model within a certain
region of parameter space. However because of the
different assumed symmetries of the fields, the hedgehog ansatz excludes
this region. Hence, it is not surprising that I
find a new solution, one 
which is discretely self-similar~(DSS)
with echoing period $\Delta=0.46$, much smaller
than that for either the original DSS~($\Delta=3.44$) or
that of the Einstein Yang-Mills~(EYM) model ($\Delta=0.74$)~\cite{matt_eym}.

An ansatz similar to this one (an $SU(2)$ valued hedgehog ansatz) was 
used to study the collapse of Skyrmions~\cite{bizon1,bizon2,bizon3}.
Both these models allow the consideration of non-spherically symmetric
initial data, albeit quite special data, within the comparatively
simple framework of one spatial dimension, namely a radius $r$.

In Section~\ref{sec:model}, I present the most general form of
the model under consideration here. First, I choose a triplet
scalar field $\Phi^a$ accompanied by the usual symmetry-breaking
potential with vacuum value $|\Phi^a|=\eta$ and coupling $\lambda$.
For certain evolutions discussed later, I also introduce
a free, massless scalar field $\psi(r,t)$ which couples to $\Phi^a$
only through gravity.

With the most general form of the equations
of motion,  Section~\ref{sec:results} then discusses various
aspects of regions of the parameter space. In particular,
Section~\ref{sec:newdss} presents a new critical solution obtained
with the hedgehog ansatz, and Section~\ref{sec:addpsi} discusses
the stability of this solution with respect to the addition of $\psi$,
a free scalar field. Section~\ref{sec:inside} addresses 
the results obtained in the interior of a monopole. I then conclude
in Section~\ref{sec:conclusion}.

%%%%%%%%%%%%%%%%%%%%%%%%%%%%%%%%%%%%%%%%%%%%%%%%%%%%%%%%%%%%%%%%
\section{The Model}
\label{sec:model}
%%%%%%%%%%%%%%%%%%%%%%%%%%%%%%%%%%%%%%%%%%%%%%%%%%%%%%%%%%%%%%%%
Letting $\Phi^a$ represent an $SO(3)$-valued triplet scalar field, and
$\psi$ a single-valued, free, massless scalar field,
the Lagrangian for the model is
\be
L =
     - \frac{1}{2} \psi_{;\mu} \psi^{;\mu}
     - \frac{1}{2} \Phi^a{}^{;\mu} \Phi^a{}_{;\mu}
     - \frac{1}{4}\lambda \left[ \left( \Phi^a \right)^2 - \eta^2 \right]^2,
\label{eq:lagrangian}
\ee
where $\lambda$ is the coupling to the symmetry-breaking potential,
and $\eta$ is the scale of symmetry-breaking.
The stress-energy tensor is then
\beq
T_{\mu \nu}
 & = &
   \psi_{;\mu} \psi_{;\nu}
 - \frac{1}{2} g_{\mu \nu}
               \psi^{;\rho} \psi_{;\rho} \nonumber \\
 &  & + \Phi^a{}_{;\mu} \Phi^a{}_{;\nu}
 - \frac{g_{\mu \nu}}{2} \left(
           \Phi^a{}^{;\rho} \Phi^a{}_{;\rho}
         + \frac{1}{2}\lambda \left[ \lb \Phi^a \rb^2 - \eta^2\right]^2 \right)
\eeq
and the equations of motion for the matter fields are
\beq
\Box \psi & = & 0 \\
\Box \Phi^a
& = & \lambda \Phi^a \left[ \lb \Phi^a\rb^2 - \eta^2 \right].
\eeq

The hedgehog ansatz is assumed for the triplet scalar field
\be
\Phi^a = f(r,t)~\hat r
       =  f(r,t) \left( \begin{array}{d}
                            \sin \theta \cos \varphi \\
                            \sin \theta \sin \varphi \\
                            \cos \theta
                       \end{array} \right),
\ee
such that the triplet field is defined to be a vector in
internal space of magnitude $f(r,t)$ pointing everywhere in the
radial direction, $\hat r$. None of the fields $\Phi^a$ is a function
only of $r$ and $t$, but the magnitude $f$ is. Thus, while highly
symmetric, the model is not spherically symmetric in the strictest
sense.
With this hedgehog ansatz,
a monopole of unit charge is obtained by enforcing the boundary condition
$f(\infty,t)=\eta$. When $f(\infty,t)=0$, no monopole charge exists.

I choose the spherically symmetric metric
\be
ds^2 = - \alpha^2(r,t)~dt^2
       + a^2(r,t)~dr^2
       + r^2~d\Omega^2,
\label{eq:metric}
\ee
and introduce auxiliary variables
\beq
\Pi_\psi \equiv \frac{a}{\alpha} \dot \psi, & ~~~~ & \Phi_\psi \equiv \psi',\\
\Pi_f    \equiv \frac{a}{\alpha} \dot f,    & ~~~~ & \Phi_f    \equiv f',   
\eeq
where an overdot denotes $\partial/\partial t$
and a prime denotes $\partial/\partial r$.
The full system of equations then becomes
\beq
\dot \psi & = & \frac{\alpha}{a} \Pi_\psi 
~~~~~~~~~~~~~~~~
%%%%
\dot \Phi_\psi = \lb \frac{\alpha}{a} \Pi_\psi \rb' 
\label{eq:eom1} \\
%%%%
\dot \Pi_\psi
& = & \frac{1}{r^2} \lb
          \frac{\alpha}{a} r^2 \Phi_\psi \rb'\\
%%%%
\dot f & = & \frac{\alpha}{a} \Pi_f 
~~~~~~~~~~~~~~~~
%%%%
\dot \Phi_f = \lb \frac{\alpha}{a} \Pi_f \rb' \\
%%%%
\dot \Pi_f
& = & \frac{1}{r^2} \lb
          \frac{\alpha}{a} r^2 \Phi_f \rb'
%-\alpha a \lambda f \left( f^2 -\eta^2\right)
-\alpha a \lambda f \Bigl( f^2 -\eta^2\Bigr)
-\frac{2\alpha a f}{r^2}\\
%%%%
\frac{a'}{a} & = & \frac{1 - a^2}{2r}
+ 2 \pi r \Bigl(  \Pi_f^2 + \Phi_f^2  + \Pi_\psi^2 + \Phi_\psi^2 \Bigr)
\nonumber \\
& &
+ 4 \pi a^2 \left[   \frac{f^2}{r}
                          + \frac{r}{4}\lambda \bigl( f^2 - \eta^2 \bigr)^2
                     \right]\\
%%%%
\frac{\alpha'}{\alpha}  & = & 
\frac{a^2-1}{2r}
+2 \pi r \Bigl(  \Pi_f^2 + \Phi_f^2  + \Pi_\psi^2 + \Phi_\psi^2 \Bigr) 
\nonumber \\
& &
-4 \pi a^2  \left[
                   \frac{f^2}{r}
                 + \frac{r}{4} \lambda \bigl( f^2 - \eta^2 \bigr)^2
                            \right]\\
%%%%
\dot a & = & 4 \pi r \alpha \Bigl( \Phi_f \Pi_f  + \Phi_\psi \Pi_\psi \Bigr) .
\label{eq:eom2}
\eeq
In addition to these equations of motion, appropriate boundary
conditions are needed. At the origin of coordinates, regularity
is enforced with the conditions
\beq
\Phi_\psi(0,t)  & = &0  \cr
f(0,t)          & = &0   ~~~~~~~~~~  \Phi_f'(0,t) = 0 \cr
a(0,t)          & = &1   ~~~~~~~~~~~\hspace{0.02in}  a'(0,t)      = 0 \\
\alpha'(0,t)    & = &0. \nonumber
\eeq
Further, a mass aspect ratio $m(r,t)$ is defined by associating
the $g_{rr}$ metric components of~(\ref{eq:metric}) and Schwarzschild
\be
a^2 = \left( 1-\frac{2m}{r}\right)^{-1}.
\ee
The quantity $m(r,t)$ then describes the amount of mass contained
within the radius $r$ at time $t$.

At large radius, an outgoing radiation condition is applied to
the scalar field $\psi$. For non-vanishing potential, the field
$f(r,t)$ has an effective mass which makes finding a good outer boundary
condition difficult. While the use of an outgoing
radiation boundary condition worked well in some situations,
in general the value of $f$ at the outer boundary was simply held
%constant which worked sufficiently whether or not a monopole was present.
constant which sufficed whether or not a monopole was present.
Reflection off this boundary was often significant so
that the maximum radius was kept relatively large. This prevented
reflections from affecting the dynamics near the origin.

%%%%%%%%%%%%%%%%%%%%%%%%%%%%%%%%%%%%%%%%%%%%%%%%%%%%%%%%%%%%%%%%
\section{Results}
\label{sec:results}
%%%%%%%%%%%%%%%%%%%%%%%%%%%%%%%%%%%%%%%%%%%%%%%%%%%%%%%%%%%%%%%%

The field equations (\ref{eq:eom1}-\ref{eq:eom2}) are then integrated with
a second-order accurate, Crank-Nicholson finite
difference code. The generated solutions are found to be
second-order convergent, stable, and mass conserving. Because either
$\lambda$ or $\eta$ can be rescaled away, I choose either $\lambda=0.1$
or $\lambda=0$
for all that follows. As almost all other studies of Type~II critical
behavior have, I have made use of an adaptive mesh. In particular, I
incorporated finite difference
approximations to the equations~(\ref{eq:eom1}-\ref{eq:eom2})
into the adaptive infrastructure developed by Choptuik~\cite{matt}.

%%%%%%%%%%%%%%%%%%%%%%%%%%%%%%%%%%%%%%%%%%%%%%%%%%%%%%%%%%%%%%%%
\subsection{A New DSS Critical Solution}
\label{sec:newdss}
%%%%%%%%%%%%%%%%%%%%%%%%%%%%%%%%%%%%%%%%%%%%%%%%%%%%%%%%%%%%%%%%
As the Lagrangian presented in~(\ref{eq:lagrangian}) is quite general,
I first discuss results obtained with vanishing scalar field $\psi$.
Specifically, if $\psi(r,t)$ is initialized initially to zero with
vanishing time derivative it will remain vanishing. This simplification
allows the study of the collapse of the triplet field alone.

I also postpone discussion of collapse within a monopole until
Section~\ref{sec:inside}. Hence,
in this section 
no monopole is present and $f(\infty,t)=0$.

\begin{figure}
\epsfxsize=8cm
\centerline{\epsffile{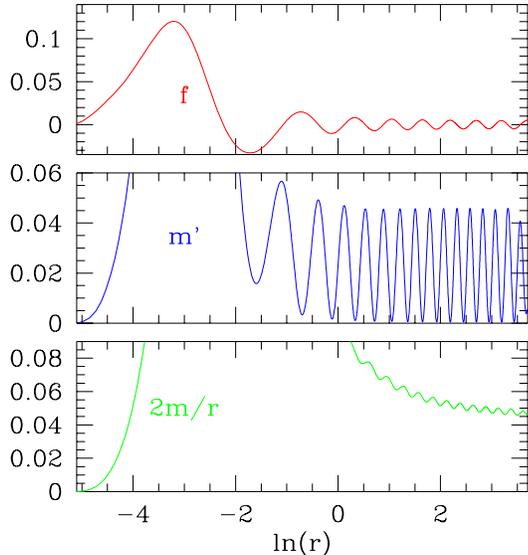}}
\caption{
         Marginally sub-critical solution at late time for $\eta=0$. The top
         frame displays the hedgehog profile $f$, the middle
         frame shows the quantity $dm/dr$ explained in the text,
         and the last frame shows the value of $2m/r$.
         }
\label{fig:crit}
\end{figure}

%I find that as the threshold of black hole formation is approached,
I find as the threshold of black hole formation is approached,
a discretely self-similar solution is found with period $\Delta = 0.46$.
A solution which approaches this precisely critical solution
 is shown in Fig.~\ref{fig:crit}. In the limit of precise criticality,
the solution is expected to exhibit an infinite number of echoes
and to be singular at the origin at the time of collapse, occurring
at a finite proper time.

\begin{figure}
\epsfxsize=9cm
\centerline{\epsffile{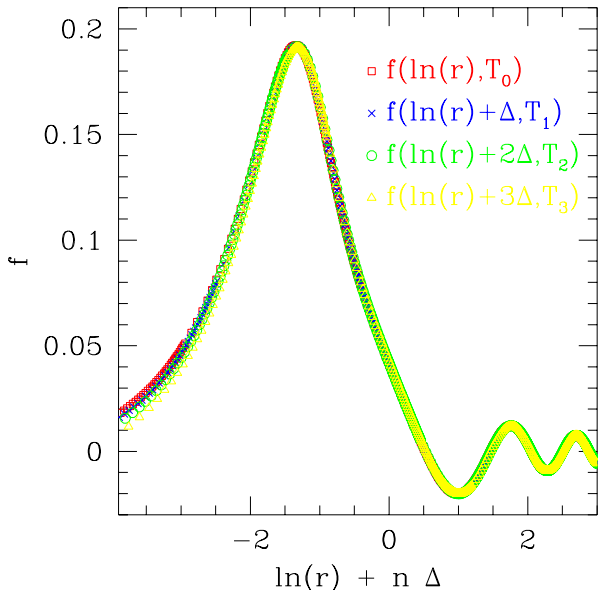}}
\caption{
         Demonstration of scale-periodicity of the new critical
         solution with $\Delta=0.46$. Here, the field $f$
         is shown at four different times for a marginally
         sub-critical solution. The times and the value of
         $\Delta$ were chosen to minimize the difference
         $f(\ln(r),T_0)-f(\ln(r)+\Delta,T_1)$.
         Independently, the value of $\Delta$ was
         also computed
         by $\Delta = \ln((T^*-T_n)/(T^*-T_{n+1}))$ 
         where $T$ represents the central proper time and $T^*$
         the estimated critical time of collapse.
         The other two profiles
         are then shown shifted in $\ln(r)$ by $n\Delta$.
         The agreement demonstrates the
         scale periodicity.
         }
\label{fig:echoing}
\end{figure}

\begin{figure}
\epsfxsize=9cm
\centerline{\epsffile{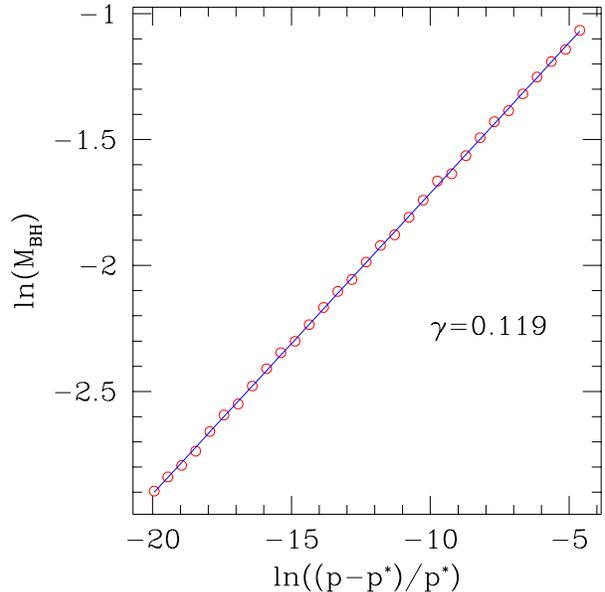}}
\caption{
         Demonstration of black hole mass scaling for $\eta=0$.
         The circles indicate the mass of black holes formed
         versus a normalized distance from criticality.
         The line indicates a
         least-squares fit for slope $\gamma=0.119$ where
         $M_{\rm BH} \propto |p-p^*|^\gamma$.
         }
\label{fig:mass}
\end{figure}

The discrete nature of the self-similarity is demonstrated in
Fig.~\ref{fig:echoing}. This solution also exhibits the
usual mass scaling relationship in the super-critical regime,
demonstrated in Fig.~\ref{fig:mass}.

Numerical results indicate that variations of $\eta$
in the interval $[0,0.5]$ do not alter the nature of the
critical solution. In other words, the discrete nature
remains with $\Delta=0.46$.
The fact that the triplet field's value for large radius vanishes
while the true vacuum lies at $\eta$ implies that
the mass of the spacetime is divergent with increasing $r$.
Hence, the profiles of $m'$ and $2m/r$ as shown in Fig.~\ref{fig:crit}
are not the same as when
$\eta=0$, differing for large $r$. However, the phenomena associated
with approaching criticality occur for ever decreasing $r$,
and in this regime, the solution remains discretely self-similar
with period $\Delta=0.46$.
This result is consistent with the potential
not playing a role in the kinetic-energy dominated critical behavior.
In the language of~\cite{carsten_rev}, the potential becomes
``asymptotically irrelevant.''

This divergence of the mass is not as problematic as it might
initially appear because the numerical grid is artificially
limited to some finite value of $r$. The mass is therefore
limited, but this cut-off should not affect results for large $r$ because
of the distance needed for any effect to travel from the outer
boundary. By comparing otherwise identical evolutions with 
different maximum radius, no such effect was seen.

It is interesting to note however that when $\eta \ne 0$
and $f(\infty,t)=0$, the
value of $2m/r$ was seen to be divergent.
This situation
represents an infinite mass spacetime  (essentially an infinity
of false vacuum energy) and should
have a horizon at some radius where $2m/r\rightarrow 1$.
Hence, it would
seem that the critical behavior
observed here is occurring within this horizon.

%%%%%%%%%%%%%%%%%%%%%%%%%%%%%%%%%%%%%%%%%%%%%%%%%%%%%%%%%%%%%%%%
\subsection{Addition of a regular scalar field}
\label{sec:addpsi}
%%%%%%%%%%%%%%%%%%%%%%%%%%%%%%%%%%%%%%%%%%%%%%%%%%%%%%%%%%%%%%%%
An interesting result found by Choptuik in his study of collapse
of a Yang-Mills field, is that the presence of even an
infinitesimal amount of scalar field
added to the Einstein Yang-Mills model drives the solution
to the scalar field's critical solution at criticality~\cite{matt_per}.
Thus, the Yang-Mills field is unstable to the
growth of the scalar field at criticality.

To test for a similar effect here, I allow $\psi$ to be non-vanishing
and look for the critical solution. Specifically, allowing
the fields $f(r,t)$ and $\psi(r,t)$ to be arbitrary Gaussian
pulses, I can ``mix'' the amplitudes with a parameter $\sigma$
such that for $\sigma=0$, the field $\psi$ vanishes and for
$\sigma=1$, the field $f$ vanishes.

\begin{figure}
\epsfxsize=9cm
\centerline{\epsffile{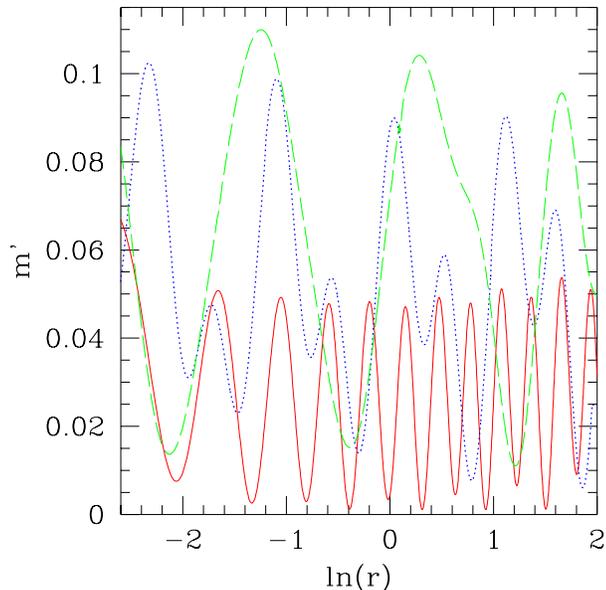}}
\caption{
         %Transition of critical solution in $\sigma$ from
         Transition of critical solution from
         %new DSS ($\Delta=0.46$) to the original DSS ($\Delta=3.44$).
         $\Delta=0.46$ to $\Delta=3.44$.
         A series of near-critical solutions are shown for $\sigma=0.4$ (solid),
         $\sigma=0.5$ (dotted), and $\sigma=0.6$ (dashed). The critical
         solution for small $\sigma$ is the new DSS, and as $\sigma$ increases,
         the critical solution becomes the original DSS. In the transition
         region, near-critical solutions appear to demonstrate structure from
         both critical solutions.
         }
\label{fig:sigma}
\end{figure}

For the EYM case, any initial data for which
$\sigma\ne0$ results in the scalar field critical solution
being found~\cite{matt_per}. Here, no such sensitivity is observed.
Instead, within some finite range of $\sigma$ the critical
solution is observed to transition between the DSS of the
scalar field ($\Delta=3.44$) and the new DSS presented here ($\Delta=0.46$).

As $\sigma$ is varied across the transition from the original DSS to 
the new one,
the fields $f$ and $\psi$ exchange dominance, both echoing at their
respective value of $\Delta$. In the transition region, the near-critical solution,
as represented by the field $m'(r,t)$, displays this dual echoing and hence
has a non-trivial structure.
This transition is shown in Fig.~\ref{fig:sigma}.

%%%%%%%%%%%%%%%%%%%%%%%%%%%%%%%%%%%%%%%%%%%%%%%%%%%%%%%%%%%%%%%%
\subsection{Critical Phenomena Inside a Monopole}
\label{sec:inside}
%%%%%%%%%%%%%%%%%%%%%%%%%%%%%%%%%%%%%%%%%%%%%%%%%%%%%%%%%%%%%%%%
In contrast to the cases presented above in which $f(\infty,t)=0$,
I consider initial data which have unit monopole charge as indicated
by $f(r,t)$ asymptoting to $\eta$.
Here, collapse within a monopole 
is modeled by first solving for the static monopole and then
adding an in-going Gaussian perturbation to the profile $f(r,t)$.
This solution has the benefit of not being dynamic near the outer
boundary. The Gaussian pulse is then parameterized by an amplitude
$p$, and a search is conducted in $p$ for a near critical solution.

\begin{figure}
\epsfxsize=8cm
\centerline{\epsffile{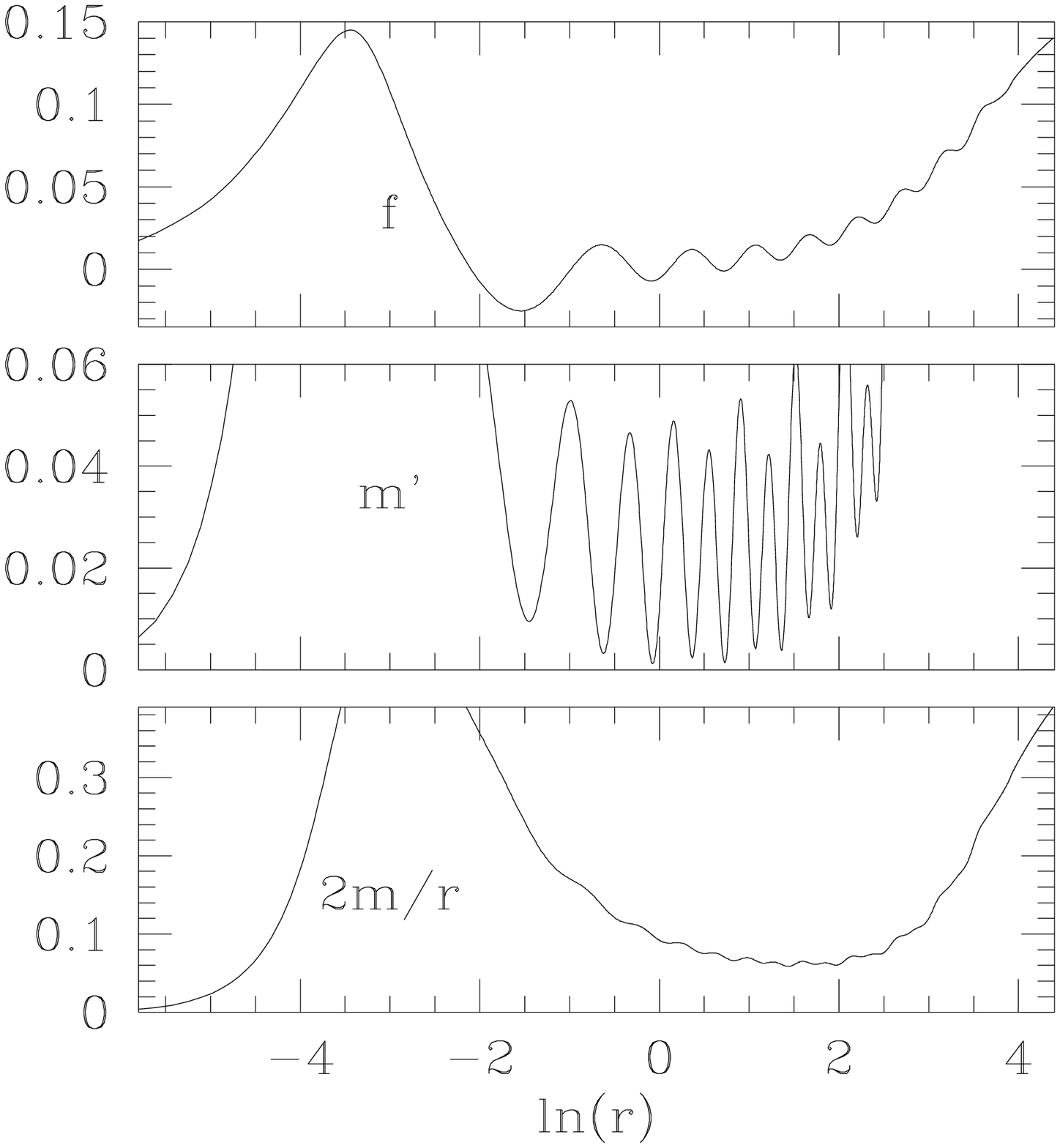}}
\caption{
         Near-critical solution obtained within a monopole for $\eta=0.15$.
         Notice that the field $f$ asymptotes to $\eta$ for large $r$
         indicating the unit charge of the configuration.
         The discrete nature with $\Delta=0.46$ is observed to be the same
         as in the vanishing potential case.
         In comparison to the $\eta=0$ case (see Fig.~\ref{fig:crit}),
         in the scaling regime (small $r$) the echoing behavior
         is evident, and
         in the large $r$ regime the solutions are different.
         }
\label{fig:mono}
\end{figure}

As stated, these evolutions should have the same critical solution
as those lacking the potential. This is verified numerically 
for $\eta \in [0,0.2]$.
For $\eta$ large ($\eta \agt 0.20$), no horizon-less, static
solutions exist~\cite{steve_future} and hence the code is able
to model these monopoles only within their horizons.
A near-critical solution for $\eta=0.15$ is shown in Fig.~\ref{fig:mono}.

That static monopoles exist in this model naturally leads one to question
whether any Type~I critical phenomena exists. However, evolutions
appear to indicate that the static monopoles with no horizon are
not unstable to collapse to a black hole.
Hence, no such Type~I critical phenomena has been observed.

%%%%%%%%%%%%%%%%%%%%%%%%%%%%%%%%%%%%%%%%%%%%%%%%%%%%%%%%%%%%%%%%
\section{Conclusion}
\label{sec:conclusion}
%%%%%%%%%%%%%%%%%%%%%%%%%%%%%%%%%%%%%%%%%%%%%%%%%%%%%%%%%%%%%%%%

A new critical solution is found by assuming a hedgehog ansatz 
for the triplet scalar field. The critical solution exhibits
discrete self-similarity with echoing period $\Delta=0.46$.
The solution also exhibits power-law mass scaling with exponent
$\gamma=0.119$.
Results verify that the potential does not affect the critical
behavior as expected.

Furthermore, the threshold of black hole formation is examined within
a global monopole. Though the potential is expected to play no
role in Type~II critical behavior, it does provide for
the existence of static solutions, namely global monopoles.
These static solutions then provide for the possibility that Type~I behavior 
is present. The static solutions would have to be unstable
to black hole formation for this Type~I behavior to exist.

However, evolutions
do not show any instability of the monopole to 
collapse, and no Type~I critical behavior is
seen.
A linear perturbation analysis should
be sufficient to settle the question of their stability.
If the static solutions
are indeed stable, then no Type~I behavior would be expected.

Finally, as discussed in~\cite{liebling} symmetries of
the initial data can determine which critical solution is attracting.
For the spherically symmetric complex scalar field,
in addition to Choptuik's original critical solution, another
critical solution exhibiting a continuous self-similarity~(CSS)
exists~\cite{hirschmann}. An arbitrary family of initial data will
find Choptuik's solution, not the CSS. However, certain types
of initial data which maximize the $U(1)$ charge density are
unable to shed their charge density and find instead the CSS.

Here, instead of a complex scalar field (i.e. a doublet field),
the single scalar field is extended to a triplet field. Within
the triplet scalar field model before making any assumptions on
the triplet, at least two critical solutions exist, both DSS.
The appropriate question is then will arbitrary families of initial
data find Choptuik's original solution, the one presented here,
or yet some other solution. To answer this with dynamical evolutions
would require a general, three-dimensional collapse code.

As a first  step towards the goal of answering this question,
Martin-Garcia and Gundlach~\cite{gundlach} have shown that
non-spherical perturbations to Choptuik's DSS decay and hence
that at least some families of non-symmetric initial data will
find Choptuik's DSS. However
this work considered only a single scalar field, and does not
address the case where a triplet is present.

The question of the relative stability of the original DSS to this
new DSS essentially reduces to determining whether a family
of generic, three-dimensional initial data for a triplet scalar
field will shed its ``in-phase'' component to become the hedgehog
DSS, or instead shed all its ``out-of-phase'' (i.e. non-zero
monopole charge density) component to become the scalar field
DSS.

%%%%%%%%%%%%%%%%%%%%%%%%%%%%%%%%%%%%%%%%%%%%%%%%%%%%%%%%%%%%%%%%
\section*{Acknowledgments}
\label{sec:ack}
%%%%%%%%%%%%%%%%%%%%%%%%%%%%%%%%%%%%%%%%%%%%%%%%%%%%%%%%%%%%%%%%

I would like to thank Matthew Choptuik for providing me with his
adaptive mesh framework.
I am also thankful for the support of the Southampton College Research
\& Awards Committee.

%**********************************************************************

\end{document}